\documentclass[11pt]{article}

\usepackage[margin=1in]{geometry}
\usepackage{setspace}
\usepackage{amsmath, amssymb}
\usepackage{graphicx}
\usepackage{natbib}
\usepackage{hyperref}

\title{
Incentives, Equilibria, and the Limits of Healthcare AI\\
\large A Game-Theoretic Perspective
}

\author{
Ari Ercole\textsuperscript{1,2}\\[0.5em]
\small \textsuperscript{1}Cambridge Centre for AI in Medicine, University of Cambridge, UK\\
\small \textsuperscript{2}Magdalene College, University of Cambridge, UK\\
\small ae105@cam.ac.uk\\
\small ORCID: 0000-0001-8350-8093
}

\date{}

\begin{document}

\maketitle

\begin{abstract}
Artificial intelligence (AI) is widely promoted as a response to capacity and productivity pressures in healthcare systems. Despite demonstrable local efficiency gains, system‑level transformation has been limited. This paper proposes a simple analytic lens to explain this gap.

Using a stylised coordination problem drawn from inpatient capacity management, three archetypal forms of AI deployment are described: effort‑reducing technologies, observability‑oriented systems, and interventions that alter underlying incentive structures. Effort reduction and observability may improve performance within existing patterns of behaviour but do not, in general, change which actions are individually rational. As a result, such interventions are typically absorbed into existing equilibria.

By contrast, interventions that modify how local actions map to downstream consequences—by redistributing or bounding local risk—can change stable system behaviour. These mechanism‑level interventions differ not in technical sophistication but in their interaction with institutional incentives.

The analysis suggests that expectations of system‑level gains from AI should be conditioned on whether a deployment changes incentives rather than optimising tasks or information flows alone. For healthcare organisations and policymakers, this has practical implications for procurement, governance, and evaluation of digital technologies.
\end{abstract}

\section{Introduction}

Artificial intelligence (AI) is increasingly presented as a potential general-purpose response to the capacity, cost, and productivity pressures facing healthcare systems. Advances in machine learning, natural language processing, and generative models have enabled AI systems to operate across a wide range of clinical and administrative tasks, leading to expectations that their diffusion will deliver not only local efficiency gains but system-level transformation. Despite this, realised productivity gains in healthcare are not a given; while localised productivity gains are demonstrable in tightly controlled settings \citep{abramoff2023}, evidence from randomised trials remains largely confined to single-centre settings 
with surrogate endpoints, and demonstrable gains have not translated reliably to system-level outcomes \citep{han2024}; broader adoption across healthcare systems remains limited \citep{sahni2023}.

A common explanation for this gap is that AI has not yet been adopted at sufficient scale or technical maturity and indeed this is a relatively new technology in the healthcare setting. An alternative explanation, explored in this paper, is that the limiting factor is not individual task performance but coordination. Healthcare delivery requires the continuous coordination of work across professional, organisational, and temporal boundaries under uncertainty. These coordination requirements impose transaction costs \citep{north1991institutions,north1984transaction} that shape behaviour even when technical capacity exists. When the local costs of coordination are borne by individual teams while the benefits are shared across the system, stable patterns of behaviour can emerge that are individually rational but collectively inefficient.

This perspective suggests that the impact of AI depends critically on how it interacts with existing incentive structures. Technologies that reduce the effort required to perform tasks or that improve the visibility of system state may improve local efficiency, but they do not necessarily change which actions are individually rational. As a result, such interventions can be absorbed into existing equilibria rather than displacing them. Understanding when AI can change system-level behaviour therefore requires an explicit account of incentives and equilibrium selection, rather than a focus on technical capability alone.

Such considerations motivate a game-theoretic approach to healthcare coordination. 
This paper applies established tools from mechanism design and coordination game 
theory \citep{roth2002engineer} to the novel setting of healthcare AI deployment. To develop this account, the paper re-applies established game-theoretic reasoning to contemporary healthcare AI deployment. Rather than proposing a new model, I use a minimal formalisation to analyse how different classes of AI intervention modify a simple coordination game. Three archetypal forms of AI deployment are introduced as an analytic lens: effort-reducing interventions that lower task costs, observability-oriented interventions that change what is visible or actionable within the system, and mechanism-level interventions that alter how local actions map to downstream consequences.

My core claim is that these archetypes differ not in technical sophistication but in how they modify the underlying game. Effort reduction and observability can improve performance within a given equilibrium, but only mechanism-level interventions can change which equilibria are stable. The analysis is grounded in an operational example drawn from inpatient capacity management, but the argument is intended to generalise to other coordination--constrained settings in healthcare. The implications are practical: expectations of system--level transformation from AI should be conditioned on whether a proposed intervention changes incentives rather than optimising tasks within an unchanged structure.

\section{Operational example: inpatient capacity signalling}

Inpatient capacity management provides a concrete (and important) illustration of a coordination problem. Suppose that individual wards repeatedly face a choice between exposing/creating available capacity to the wider system or buffering it locally. Exposing capacity can improve overall patient flow, but it can also trigger admissions and workload that are borne locally. Buffering capacity protects local workload and safety margins, but can contribute to congestion elsewhere. The example does not rely on assumptions about motivation or compliance. It follows from how costs and benefits are distributed. Box 1 summarises the archetypes.

\begin{center}
\fbox{
\begin{minipage}{0.95\linewidth}

\textbf{Baseline (no AI).}
Local exposure of capacity carries higher local cost than buffering, while unilateral exposure yields little marginal system benefit. Buffering is therefore a best response to buffering by others, and $(B,\dots,B)$ is a Nash equilibrium.

\medskip
\textbf{Effort-reducing AI.}
Effort-reducing interventions lower the cost of acting but do not alter the relative disadvantage of exposure.

\medskip
\textbf{Observability-oriented AI.}
Observability-oriented interventions add an expected organisational consequence to buffering when it is detected.

\medskip
\textbf{Mechanism-level AI.}
Mechanism-level interventions bound or redistribute the local downside of exposing capacity, making cooperation individually rational.
\end{minipage}
}
\end{center}

\subsection{Baseline game}

Let wards be indexed by \( i = 1,\dots,N \). Each ward chooses an action
\[
a_i \in \{E,B\},
\]
where \(E\) denotes exposing capacity and \(B\) denotes buffering capacity. Payoffs are given by
\[
u_i(a_i,a_{-i}) = b(a_i,a_{-i}) - c_i(a_i),
\]
where \(b(\cdot)\) captures benefits from improved system flow and \(c_i(\cdot)\) captures local workload and risk costs.

The structural asymmetry is that exposure is locally riskier than buffering:
\[
c_i(E) > c_i(B).
\]
When other wards buffer, unilateral exposure yields little marginal system benefit:
\[
b(E,B_{-i}) \approx b(B,B_{-i}).
\]
Under these conditions, the best-response inequality holds:
\[
u_i(B,B_{-i}) > u_i(E,B_{-i}),
\]
so \(B\) is a best response to \(B_{-i}\). Therefore \((B,\ldots,B)\) is a Nash equilibrium.

A cooperative outcome \((E,\ldots,E)\) may be socially preferable, but it is not an equilibrium unless the local disadvantage of exposure is removed or reversed.

\subsection{Archetype-specific modifications}

Each archetype is treated as a modification to the baseline game. For each, the question is whether the modification changes the best-response comparison between \(E\) and \(B\).

\subsubsection{Effort-reducing AI}
Many contemporary AI deployments focus on reducing the effort required to perform 
coordination-related tasks. Representative examples include ambient voice documentation, 
large language model assisted discharge summary drafting, AI-generated referral letters, 
and clinical inbox triage with draft response generation. Such technologies can be highly 
effective at reducing friction in coordination work, but they typically change only the 
time cost of acting.

In game-theoretic terms, effort reduction modifies the cost term:
\begin{equation}
    c_i(a_i) \rightarrow c_i(a_i) - \Delta(a_i),
\end{equation}
where $\Delta(a_i) \geq 0$ is the reduction in effort cost associated with action 
$a_i$. The best-response comparison to others buffering becomes:
\begin{equation}
    u_i(B, B_{-i}) - \Delta(B) \geq u_i(E, B_{-i}) - \Delta(E).
\end{equation}
Unless effort reduction disproportionately lowers the cost of exposure relative to 
buffering, the inequality remains true and $(B,\ldots,B)$ remains a Nash equilibrium. 
Local efficiency can improve without a change in system-level behaviour.

\subsubsection{Observability-oriented AI}
A second class of AI technologies focuses on making system state more visible or 
predictable, for example models forecasting discharge timing or congestion risk, 
analytics that make delays more visible, or alerting systems that surface deviations 
from expected patterns. These tools change what is visible or actionable within 
organisational processes, but they do not directly reduce the local workload or risk 
associated with exposure.

In the game, observability adds an expected organisational consequence to buffering 
when buffering is detected and acted upon:
\begin{equation}
    u_i(B, a_{-i}) \rightarrow u_i(B, a_{-i}) - p(a_{-i})F,
\end{equation}
where $p(a_{-i}) \in [0,1]$ is the probability that buffering is noticed and acted 
upon given others' behaviour, and $F \geq 0$ is the expected consequence. Note that 
$p$ is treated here as exogenous; in practice, strategic actors may adjust behaviour 
to reduce detectability, which would weaken this route to equilibrium change.

Buffering remains a best response whenever:
\begin{equation}
    u_i(B, B_{-i}) - p(B_{-i})F \geq u_i(E, B_{-i}).
\end{equation}
Only if the expected consequence outweighs the baseline advantage of buffering does 
a different equilibrium become possible. This route to equilibrium change operates 
through added expected cost rather than through reducing the underlying local risk 
of exposure.

\subsubsection{Mechanism-level AI}

A third, qualitatively different possible class of AI intervention, not yet prevalent in healthcare, operates at the level of mechanism design. Rather than reducing effort or increasing visibility, such systems restructure how local actions map to local consequences. The defining feature is institutional rather than technical: the system, rather than the individual ward, absorbs a defined share of the downside risk that makes cooperation individually irrational. Examples include coupling early capacity disclosure to pre-committed staffing redeployment, where the commitment is made before the admission occurs and not contingent on subsequent negotiation, or transferring responsibility for downstream sequencing and escalation to a central function once capacity is revealed.

In the game, let $\kappa$ denote a mechanism that bounds or redistributes the local cost of the exposure action $E$. This is represented as:
\begin{equation}
    c_i(E) \rightarrow c_i(E \mid \kappa),
\end{equation}
where $c_i(E \mid \kappa) < c_i(E)$. Exposure becomes a best response to buffering when:
\begin{equation}
    u_i(E, B_{-i} \mid \kappa) \geq u_i(B, B_{-i}),
\end{equation}
equivalently:
\begin{equation}
    b(E, B_{-i}) - c_i(E \mid \kappa) \geq b(B, B_{-i}) - c_i(B).
\end{equation}
If this condition holds for all wards, $(E,\ldots,E)$ becomes a Nash equilibrium. Unlike observability-oriented interventions, this change does not rely on sustained monitoring or enforcement. It follows directly from the incentive structure, because revealing capacity no longer exposes the ward to unbounded local risk.

\section{Discussion}

The operational example illustrates a general point about healthcare AI: the binding constraint on system-level change is often not task performance or information availability, but the incentive structure governing coordination. In the baseline game, buffering capacity is individually rational even when collective exposure would improve system performance. The three archetypes differ in how they interact with this structure, and therefore in whether they can plausibly shift equilibrium behaviour.

Effort-reducing AI lowers the cost of acting but leaves the relative disadvantage of exposure intact. Observability-oriented AI changes what is visible and actionable, but typically does so by adding expected consequences rather than by removing underlying local risk. In both cases, behaviour may change at the margin, but the equilibrium itself remains stable unless the intervention is strong enough to reverse the best-response inequality. This helps explain why many AI deployments deliver real local value while failing to produce measurable system-level transformation.

Mechanism-level AI differs because it changes the game rather than optimising play within it\citep{roth2002engineer}. By bounding or redistributing the local downside of cooperation, these interventions can make exposing capacity individually rational. When this condition holds across actors, cooperative behaviour becomes a Nash equilibrium rather than a fragile coordination outcome. Importantly, this does not require automation of clinical decisions or coercive enforcement. It requires institutional commitment to absorb coordination risk at the system level rather than returning it to local units.

The analysis does not depend on wards being symmetric. In practice, capacity decisions are often asymmetric, with some wards facing higher baseline costs from exposure due to staffing levels or case mix, while others benefit more from buffering. This asymmetry does not eliminate the coordination problem. It typically sharpens it, because actors with higher exposure costs are even less willing to cooperate unless the mechanism explicitly reduces or compensates those costs.

Many real healthcare coordination problems also depart from simple two-action games. In some settings, individual units possess veto power or face threshold effects, where a single decision can block flow, delay a pathway, or trigger escalation. These situations can be represented as veto or threshold games rather than pure coordination games. The core insight remains unchanged. Effort reduction and observability may alter the ease or visibility of veto use, but they do not alter the incentive to exercise it when doing so protects local risk. Mechanism-level interventions remain the only class that can reliably change outcomes, because they change what exercising a veto entails in terms of local cost.

The static Nash analysis presented here is deliberately minimal. In practice, healthcare coordination is repeated, adaptive, and embedded in organisational learning. An evolutionary game-theoretic perspective provides a natural extension. Over time, strategies that protect local actors from unbounded risk are likely to persist and spread, even without explicit coordination failure. Conversely, mechanisms that reliably bound the downside of cooperation create conditions under which cooperative strategies can be selected and stabilised. From this perspective, AI does not need to explicitly solve coordination. It needs to change the fitness landscape in which behaviours evolve.

This framing clarifies why procurement and governance choices matter. Healthcare systems frequently adopt AI tools aligned with effort-reducing or observability-oriented archetypes because these deliver immediate and visible benefits to individual users. Mechanism-level interventions, by contrast, require system-level ownership, redesign of workflows, and explicit acceptance of redistributed risk. They are therefore harder to procure, harder to govern, and harder to evaluate using conventional metrics. The analysis here suggests that without such interventions, expectations of system-level productivity gains from AI are likely to be disappointed, regardless of technical progress.

In summary, the contribution of this paper is not to propose a new model of healthcare delivery, but to provide an analytic lens for distinguishing between AI interventions that optimise behaviour within existing equilibria and those that can change them. This distinction is orthogonal to technical sophistication and central to claims of transformation. It suggests that the critical question for healthcare leaders is not whether AI is powerful, but whether a given deployment changes incentives in a way that makes system-level cooperation individually rational.

For healthcare organisations, this distinction has practical implications. Procurement decisions that prioritise task‑level efficiency or improved visibility alone are unlikely to deliver measurable system‑level gains if underlying incentive structures are unchanged. Conversely, deployments coupled to explicit organisational commitments—such as pre‑committed resource redistribution or centralised coordination functions—are more likely to alter behaviour. Evaluation frameworks should therefore distinguish between local performance improvements and changes in system dynamics, and funding decisions should be aligned accordingly.

\section{Conclusion}

This paper argues that the system-level impact of healthcare AI depends less on technical capability than on how AI deployments interact with incentives governing coordination. Using a minimal game-theoretic framework, it distinguishes between interventions that reduce effort, those that increase observability, and those that operate at the level of mechanism design. Only the latter class can reliably change equilibrium behaviour by making cooperative action individually rational.

The analysis provides a framework to understand which AI deployments are likely to leave aggregate outcome unchanged even though they may deliver genuine local value. Optimising tasks or increasing visibility improves performance within an existing equilibrium, but may not, in general, alter which behaviours are stable. For healthcare leaders, the implication is practical: expectations of transformation, and therefore potential investment, should be conditioned on whether an intervention changes incentives, not merely workflows or information.

\bibliographystyle{plainnat}
\bibliography{references}

\end{document}